\documentclass[a4paper,twoside]{article}
\usepackage{tikz}
\usetikzlibrary{positioning,calc}
\usepackage{epsfig}
\usepackage{subcaption}
\usepackage{calc}
\usepackage{amssymb}
\usepackage{amstext}
\usepackage{amsmath}
\usepackage{amsthm}
\usepackage{multicol}
\usepackage{pslatex}
\usepackage{apalike}
\usepackage{algorithm2e}
\usepackage{booktabs}
\usepackage[bottom]{footmisc}
\usepackage{SCITEPRESS} 
\usepackage{xcolor}
\usepackage{graphicx}
\usepackage{array}
\usepackage{pgfplots}
\pgfplotsset{compat=1.18}
\usepackage{float} 
\usepackage{tabularx}
\usepackage{array}
\usepackage{titlesec}
\usepackage{url}

\titleformat{\paragraph}
  {\normalfont\normalsize\bfseries}
  {\theparagraph}{1em}{}

\titlespacing*{\paragraph}
  {0pt}{3.25ex plus 1ex minus .2ex}{1.5ex plus .2ex}

\begin{document}

\title{A Mixed-Method Empirical Study of LLM Assistance in Software Engineering Workflows \vspace{0cm}}

\author{\authorname{Pamali D. Weerasinghe\sup{1}\orcidAuthor{0009-0007-1095-9679}, Roshan N. Rajapakse\sup{1}\orcidAuthor{0000-0002-9258-5784},
Isuru Dharmadasa\sup{2}\orcidAuthor{0000-0001-9010-1056} and
Chamath Keppitiyagama\sup{1}\orcidAuthor{0000-0003-1613-5260}
}
\affiliation{\sup{1}University of Colombo School of Computing, Colombo, Sri Lanka}
\affiliation{\sup{2}University of the Sunshine Coast, Adelaide, Australia}
\email{2022cs216@stu.ucsc.cmb.ac.lk, rnr@ucsc.cmb.ac.lk, imahaganiarach@usc.edu.au, cik@ucsc.cmb.ac.lk}
\vspace{-10mm}}


\abstract{Large Language Models (LLMs) are increasingly integrated into software development workflows, yet their effects are often discussed without distinguishing between task types, developer seniority, and verification demands. This paper presents a mixed-method empirical study of LLM-assisted software engineering with first-year and fourth-year undergraduates. Phase~1 is a preliminary survey (\(N=157\)) that characterizes LLM exposure, reliance, and trust calibration among the two groups. Phase~2 is a task-based quasi-experiment with a purposive sample from both cohorts (\(n=20\)). Here, we compare AI-assisted and non-AI conditions on a structured set of software engineering tasks spanning implementation, constraint-driven algorithm selection, and architectural reasoning. We then analyze performance outcomes alongside behavioral traces captured via screen recording and a qualitative coding process.
Survey results indicate widespread LLM adoption and substantial verification effort, alongside cohort differences in perceived LLM capability for constraint-heavy scenarios. The quasi-experiment further shows that AI assistance changes workflow structure. For example, participants frequently adopt AI-first task entry, copy--transfer integration, and AI-mediated debugging, whereas non-AI workflows rely more on documentation, prior templates, and iterative trial--error refinement. Overall, our findings suggest that the benefits of LLM assistance are task-dependent and mediated by expertise and verification practices, rather than by generation speed alone.}

\onecolumn \maketitle \normalsize \setcounter{footnote}{0} \vfill

\section{\uppercase{Introduction}}
\label{sec:introduction}

Large Language Models (LLMs) have become increasingly integrated into contemporary software development workflows, supporting a range of core activities including code generation, automated testing, documentation production, and debugging assistance \cite{haque2022think}, \cite{zheng2025towards}. As a result, today, Software Engineering (SE) practice is increasingly shaped by AI-assisted interaction patterns rather than by unaided programming alone \cite{hou2024large}. However, while prior work and industry reports suggest substantial productivity gains from LLM-based tooling or workflows, the extent and nature of these gains remain uneven across task types and developer expertise \cite{barke2023grounded}, \cite{peng2023impact}.

A key factor underlying this inconsistency is that SE work is not limited to well-specified coding problems with easily verifiable outputs. In practice, developers routinely work across multiple abstraction levels, including implementation, algorithm selection under constraints, performance-sensitive design, and trade-off analysis. These tasks often require interpreting incomplete requirements, reasoning about implicit constraints, and making decisions that remain maintainable and operationally sound beyond immediate correctness \cite{nunes2025evaluating}. In such contexts, raw generation speed is only one part of performance; the ability to validate and justify a solution is equally important.

Historically, these capabilities have been closely tied to developer experience and domain expertise. More experienced developers often rely on internalized mental models, prior exposure to system failure modes, and tacit knowledge of architectural trade-offs when operating under uncertainty \cite{latoza2006maintaining}. They are therefore better positioned to evaluate whether a solution is merely plausible or actually appropriate for a given context.

At the same time, a growing industry narrative suggests that LLMs may reduce the gap between novice and experienced developers by accelerating solution generation and lowering the barrier to implementation \cite{noy2023experimental}. This claim has important implications for staffing decisions and the distribution of engineering work. However, such claims are often made without sufficiently distinguishing between task categories or accounting for engineers' skill levels and relevant knowledge. In many cases, faster generation can be offset by additional effort spent on validating code correctness or bug fixing \cite{barke2023grounded}. Therefore, what is needed, then, is not further demonstration that LLMs can accelerate coding tasks, but a more granular investigation of how AI assistance interacts with task complexity and developer experience. 

This study investigates LLM assistance in SE workflows by focusing on \emph{developer behavior and performance across different task types}. We examine how first-year and fourth-year undergraduates engage with a structured set of \textit{Complex and Open-ended Software engineering Tasks} [COSTs] spanning implementation-level programming, constraint-driven algorithmic reasoning, and middleware/architecture design. Using a mixed-method empirical design, we combine a preliminary survey with a task-based quasi-experimental study and behavioral analysis of screen recordings to understand \emph{what} outcomes and \emph{how} participants work under AI-assisted and non-AI conditions.

Our contributions are as follows:
\begin{itemize}
    \item We present an empirical characterization of LLM usage patterns, trust calibration, verification behavior, and problem-solving strategies among first-year and fourth-year undergraduates.
    \item We design and conduct a task-based quasi-experimental study using a multi-layered SE task suite spanning implementation, algorithmic, and architectural reasoning tasks.
    \item We provide a behavioral analysis of participant workflows from screen recordings and identify distinct interaction patterns across AI-assisted and non-AI conditions.\vspace{-.5cm}
\end{itemize}

\section{\uppercase{Related Work\vspace{-.25cm}}}
\label{sec:related_work}

\subsection{LLMs for SE Tasks\vspace{-.25cm}}
Tools built on models including GPT, Claude, GitHub Copilot, and Cursor are increasingly integrated into development workflows to support software design, code generation, debugging, testing, and deployment \cite{wang2023review}, \cite{fan2023large}. Their integration into modern development environments reflects a broader shift toward AI-assisted SE. 

A growing body of empirical research has examined how prompting strategies influence LLM performance in programming tasks. Several studies report measurable improvements in coding velocity and solution quality through the use of structured prompt engineering techniques \cite{li2024humaneval}, \cite{li2024approach}, \cite{ridnik2024code}. For example, prior work emphasizes the importance of tailoring prompts to specific tools and model characteristics, particularly for widely deployed code-generation tools such as Claude and Copilot \cite{murr2023testing}. Along similar lines, \cite{wang2024selection} proposes a framework for selecting appropriate prompt engineering strategies based on code complexity, positioning complexity as a guiding metric for optimizing LLM-assisted development.

Beyond code generation, researchers have explored the application of LLMs across other phases of the SE lifecycle. Studies investigate their use in requirements engineering \cite{arora2024advancing}, \cite{hemmat2025research}, system design \cite{zhou2025using}, and software quality assurance \cite{boukhlif2024llms}, \cite{widyasari2024beyond}. Collectively, this body of work highlights the broad and expanding adoption of LLM-based tools throughout SE processes.

\subsection{Use of LLMs on COSTs}

While the above-discussed benchmarks provide standardized and reproducible evaluation settings, they also introduce potential validity concerns. Given that LLMs are trained on large-scale corpora that may include publicly available coding problems and solutions, there is a risk that performance on such datasets reflects memorization or pattern recall rather than genuine problem-solving capability \cite{carlini2022quantifying}. Consequently, to more objectively assess the benefits and limitations of LLM-assisted programming, evaluations should prioritize previously unseen tasks, as outlined in our introduction, thereby reducing the likelihood of data contamination.

Despite the recent advancements in LLM-supported development, relatively few studies have examined model performance and developer support in the context of COSTs. In a recent study, \cite{shen2026ai} investigated two groups of novice developers tasked with working on a new programming library, with only one group receiving AI assistance. The findings revealed mixed outcomes: although developers who relied heavily on AI tools demonstrated productivity improvements, these gains were accompanied by reduced learning outcomes and weaker conceptual understanding. Similarly, \cite{peng2023impact} reported that developers using GitHub Copilot completed tasks 55.5\% faster than those without AI support, underscoring the efficiency gains associated with LLM-based code generation. In another study examining collaborative work on GitHub projects, researchers compared two groups of developers ( i.e., with and without the support from Cursor), and observed significant short-term project-level productivity improvements among the AI-assisted group; however, these gains were not sustained and were associated with increased code complexity \cite{he2025does}.

\subsection{Students' Use of LLMs in SE\vspace{-.25cm}}

A growing line of research investigates how students perceive and use LLM-based tools in SE and programming courses. In a multi-university study involving five SE courses, \cite{baresi2025students} surveyed and interviewed students after structured instruction on ChatGPT and prompting strategies. Their findings identified perceived usefulness across different SE tasks, but also highlighted challenges in formulating effective prompts and adapting AI-generated artifacts to specific project needs.

Several studies have examined the impact of LLM use on student learning and performance. \cite{stoyanova2025chatgpt} conducted a two-part study with engineering undergraduates and found that while the usage of ChatGPT did not show a statistically significant association with final grades, lower-performing students tended to rely on it more heavily. In a quasi-experimental setting, \cite{sun2024would} compared ChatGPT-facilitated and self-directed programming modes and reported that students using ChatGPT exhibited more frequent debugging behaviors and encountered a higher number of error messages. However, although their coding performance improved, the difference was not statistically significant compared to their counterparts. Similarly, \cite{silva2024chatgpt} explored ChatGPT's integration in programming courses with first-semester students in Brazil, noting benefits in comprehension but raising concerns about over-reliance and diminished understanding of fundamental concepts.

At a broader scale, \cite{li2025engineering} surveyed 539 engineering students from 12 Chinese universities and found that over 40\% used LLM tools, with trust in AI-generated content emerging as a central challenge. \cite{khan2025integrating} investigated motivators and demotivators for LLM integration in Finnish SE education and proposed a roadmap for higher education institutions. These findings collectively underscore the need for pedagogical strategies that utilize the benefits of LLMs while mitigating risks to learning outcomes.

\section{Methodology\vspace{-2mm}}
\label{sec:methodology}
We use a mixed-methods design in this study with two phases: 
\begin{enumerate}
    \item Preliminary Survey: To explore developers’ preferences and experiences, profile the participants, and select suitable candidates for the experiment.
    \item Task-Based Quasi-Experiment: Measures the real-time performance, approaches and behavior patterns when working with COSTs.
\end{enumerate}

The target population consists of undergraduate students from the University of Colombo School of Computing (UCSC), Sri Lanka. Ethical approval for the study was obtained from the UCSC Ethical Review Committee (Ethics Reference No. ERC\_6\_1002).\vspace{-2mm}

\subsection{Phase 1: Preliminary Survey\vspace{-2mm}}

131 First-Year and 26 Fourth-Year undergraduates participated in the survey. The fourth-year cohort comprises students who advanced to the final year of the program on the basis of their academic performance. The survey consists of 3 sections:

The \textit{Demographics} section collected data including gender identity, academic year/performance, and the specific degree program. To establish a baseline of academic and SE technical skills, participants self-assessed their programming expertise on a scale from ``Beginner'' to ``Expert''. The Grade Point Average (GPA) was used to assess academic performance. Furthermore, to assess external experience, prior industry experience in using LLMs for SE tasks was queried.

The \textit{Familiarity with Domain Expertise \& LLMs} section assessed the participants' baseline reliance on AI tools and their engagement with AI-generated outputs. This section identified the specific contexts in which participants rely on LLMs and evaluated the reasons behind their confidence in solving problems without AI assistance. 

The \textit{Scenario-Based Comparative Questions} section captured participants' understanding of the efficiency gap between advanced learners in SE and LLMs with COSTs. This section presents four scenarios covering a key dimension of SE: handling COST, rapid prototyping under time constraints, adherence to implicit domain-specific regulations, and speed of convergence during interactive refinement.

\subsection{Phase 2: Quasi-Experiment\vspace{-2mm}}
Based on the survey responses, we selected 10 participants from each of the first-year and fourth-year cohorts and created with LLMs” and “without LLMs” groups for the task-based experiment (Table \ref{tab:tab1}, the number represents their experimental order). For the first year “with LLMs” group, participants were chosen based on their ability to effectively use LLM tools (i.e., at least one year of LLM experience) and their proficiency in one of the programming languages is either "Beginner (I need constant guidance)" or "Elementary(I know syntax but struggle with logic)".

For the control group (“without LLMs”) among first-year students, we selected participants with less than one year of LLM experience and the proficiency in one of the programming language is "Intermediate (I can build simple apps)", "Advanced (I understand architecture and optimization)","Expert (I can handle complex, industry-level tasks)". Our intention was that they would not naturally depend on AI-driven workflows and have sufficient related knowledge to complete the tasks.

Among fourth-year students, selection emphasized the iterative nature of SE. Participants were categorized based on their preferred problem-solving approach: those who favored using LLMs for rapid refinement and prompt-based iteration were assigned to the “with LLMs” group, while those who preferred manual refinement and relied primarily on human expertise were assigned to the “without LLMs” group.

During the experiment, participants were presented with a set of COSTs. Participants were not restricted by time to perform tasks. To facilitate analysis of real-time technical behaviors, screen recording software captured participants’ activity continuously throughout the tasks. The survey instruments and task specifications are available online \footnote{\url{https://bit.ly/3QhNMpa}}.

\begin{table}[t]

\begin{flushleft}
\caption{Participant Categorization}
\footnotesize
\label{tab:tab1}
\begin{tabular}{ >{\raggedright\arraybackslash}p{0.8cm} 
                 >{\raggedright\arraybackslash}p{2.7cm} 
                 >{\raggedright\arraybackslash}p{2.7cm} }
\toprule
 & \textbf{With AI Assistance} & \textbf{Without AI Assistance} \\
\midrule

First Years & [S02], [S03], [S06], [S10], [S14] & [S01], [S04], [S05], [S07], [S13] \\
\addlinespace

Fourth Years & [S08], [S09], [S15], [S16], [S18] & [S11], [S12], [S17], [S19], [S20] \\

\bottomrule
\end{tabular}
\end{flushleft}
\vspace{-2mm}
\end{table}

\subsection{Analysis}

Survey responses were analyzed using a combination of descriptive and inferential statistics. We calculated summary statistics to characterize participant demographics, LLM usage patterns, confidence, and verification behavior. For between-cohort comparisons (First-Year vs.\ Fourth-Year), we used nonparametric tests appropriate for ordinal and non-normally distributed data (e.g., Mann--Whitney U) and exact tests for categorical variables (e.g., Fisher’s exact test), and we report effect sizes alongside \(p\)-values.

To analyze participants’ experiment videos, we first transcribed their on-screen activities using Microsoft’s Azure Video Indexer\footnote{\url{https://azure.microsoft.com/en-us/products/ai-video-indexer}} and manually cross-checked with the screen recordings while adding more observations. 

We then conducted a thematic analysis \cite{braun2006using} on the scripts. First, observable behaviors were labeled inductively (e.g., task entry patterns, prompting actions, code transfer behaviors, debugging steps, verification activities) via open coding and iteratively refined via constant comparison/refinement across participants and conditions. Codes were then clustered into higher-level themes.\vspace{-4mm}

\begin{table}[t]
\centering
\footnotesize
\caption{Sample characteristics (N = 157)}
\label{tab:sample_characteristics}
\begin{tabular}{ >{\raggedright\arraybackslash}p{2.7cm} 
                 >{\raggedright\arraybackslash}p{1.5cm} 
                 >{\raggedright\arraybackslash}p{.75cm} 
                 >{\raggedright\arraybackslash}p{.75cm} }
\toprule
\textbf{Measure} & \textbf{Category} & \textbf{\textit{n}} & \textbf{\%} \\
\midrule

\textbf{Gender} 
& Male & 109 & 69.43 \\
& Female & 48 & 30.57 \\
\addlinespace

\textbf{Academic year} 
& First Year & 131 & 83.44 \\
& Fourth Year & 26 & 16.56 \\
\addlinespace

\textbf{Degree program} 
& Computer Science & 105 & 66.88 \\
& Information Systems & 52 & 33.12 \\
\addlinespace

\textbf{Industry experience} 
& Yes & 42 & 26.75 \\
& No & 115 & 73.25 \\
\addlinespace

\textbf{Years using LLMs} 
& $< 1$ year & 86 & 54.78 \\
& 1--2 years & 54 & 34.39 \\
& $> 3$ years & 17 & 10.83 \\

\bottomrule
\end{tabular}
\label{T1}
\end{table}

\section{Results and Discussion\vspace{-.25cm}}

\subsection{Preliminary Survey\vspace{-.25cm}}
157 responses were collected between January 12, 2026, and February 19, 2026, comprising 131 First-Year and 26 Fourth-Year undergraduates. 26.75\% of the respondents had worked in a role that required SE-related domain knowledge outside coursework (e.g., internship, part-time job, research assistant) (Table \ref{T1}).

\begin{table*}[t]
\centering
\small
\caption{LLM usage and verification behaviors (N = 157, 1Y = 131, 4Y = 26)}
\label{tab:llm_usage_verification}
\label{T2}
\begin{tabular}{llrrrrrr}
\toprule
\textbf{Measure} & \textbf{Category} 
& \textbf{\textit{N}} & \textbf{\%} 
& \textbf{\textit{1Y}} & \textbf{\%} 
& \textbf{\textit{4Y}} & \textbf{\%} \\
\midrule

\textbf{LLM use for coding} 
& Regularly (Daily Workflow) 
& 88 & 56.06 & 66 & 50.38 & 22 & 84.62 \\
& Occasionally (For Specific Errors) 
& 53 & 33.76 & 51 & 38.93 & 2 & 7.69 \\
& Heavily (Rely for most tasks) 
& 16 & 10.19 & 14 & 10.69 & 2 & 7.69 \\
\addlinespace

\textbf{Reliance for assignments} 
& Sometimes 
& 71 & 45.22 & 63 & 48.09 & 8 & 30.77 \\
& Often 
& 62 & 39.49 & 47 & 35.88 & 15 & 57.69 \\
& Always 
& 16 & 10.19 & 14 & 10.69 & 2 & 7.69 \\
& Rarely 
& 7 & 4.50 & 6 & 4.58 & 1 & 3.85 \\
& Never 
& 1 & 0.60 & 1 & 0.76 & 0 & 0.00 \\
\addlinespace

\textbf{Verification effort} 
& Significant (review every line) 
& 86 & 54.78 & 69 & 52.67 & 17 & 65.38 \\
& Minimal (check if it runs) 
& 60 & 38.22 & 51 & 38.93 & 9 & 34.62 \\
& None (trust it works) 
& 8 & 5.10 & 8 & 6.11 & 0 & 0.00 \\
& More than writing it myself 
& 3 & 1.90 & 3 & 2.29 & 0 & 0.00 \\

\bottomrule
\end{tabular}\vspace{-3mm}
\end{table*}

\subsubsection{LLM Exposure and Usage Patterns}

LLM use was widespread and routine (Table \ref{T2}). Over half of the full sample reported using LLMs as part of their daily coding workflow (56.06\%, $88/157$), with an additional 33.76\% ($53/157$) using them occasionally for targeted errors. 10.19\% reported heavy reliance for most coding tasks (10.19\%, $16/157$). Notably, Fourth-Year respondents showed higher daily integration (84.62\%, $22/26$) than First-Year respondents (50.38\%, $66/131$). This is consistent with the view that LLMs become increasingly embedded in routine development practice with accumulated experience.

When asked which technical domains they primarily use LLM assistance for (multi-select), respondents most frequently selected Algorithms \& Data Structures (78.34\%, $123/157$). However, substantial usage also appears in application-development domains: Frontend Development, Backend Development, and Database \& Querying (Figure~\ref{fig1}). 

\subsubsection{Verification, reliance, \& trust\vspace{-2mm}}
\paragraph{Reliance for assignments \& routine work\vspace{-2mm}} Use for assessed work was nearly universal: 94.90\% ($149/157$) reported relying on LLMs for assignments at least ``Sometimes.'' The modal response was ``Sometimes'' (45.22\%), followed by ``Often'' (39.49\%). Cohort patterns suggest different usage regimes: First-Year students show a slightly higher tail of ``Always'' reliance (10.69\%) and the only ``Never'' response (0.76\%), whereas Fourth-Year students cluster more strongly around frequent but not maximal reliance (57.69\% ``Often''). This pattern may reflect a shift from exploratory or dependency-prone use among novices to a more habitual, tool-integrated use among seniors.

\paragraph{Perceived reliability on COSTs\vspace{-2mm}}
When asked how often AI-generated code works \emph{immediately} without modification for COSTs, most respondents selected ``Often (50--80\%)'' (63.69\%, $100/157$), while 12.10\% ($19/157$) selected ``Almost Always (80--100\%),'' and 22.93\% ($36/157$) selected ``Rarely.'' Only 1.27\% ($2/157$) selected ``Never.'' This indicates that AI-generated code is perceived as frequently workable even for COSTs across respondents.

\paragraph{Verification behaviors and their relationship to perceived reliability.\vspace{-2mm}}
Verification effort was substantial in both cohorts (Table~\ref{tab:llm_usage_verification}). A majority reported ``Significant'' verification (54.78\%), and an additional 38.22\% reported at least minimal execution checks (``Minimal''). Only 5.10\% reported ``None (trust it works),'' and 1.91\% reported verification taking longer than writing the solution themselves (``More than writing it myself''). 

Importantly, even among respondents who believed AI-generated code works ``Often'' or ``Almost Always'' (75.80\% of the sample), 54.62\% still reported ``Significant'' verification. This suggests an emerging AI-generated code norm of \emph{trust-but-verify}. At the same time, perceived reliability was negatively associated with verification effort (Spearman $\rho = -0.21$, $p = 0.008$). This indicates that respondents who believe LLM outputs are more frequently correct tend to invest less time in auditing. 

\paragraph{Behavioral triage when solutions appear complex.\vspace{-2mm}}
When the AI produced a complex solution, most respondents reported asking the model to explain the code line-by-line before use (71.97\%, $113/157$). 20.38\% rewrote the logic in their own style, and 7.64\% copy-pasted and executed the solution without modification. This suggests that many students already use LLMs as \emph{interactive tutoring or comprehension aids}, and not solely as code generators. However, the copy-paste subgroup remains non-trivial, particularly given that a small fraction of respondents reports no verification.

\begin{figure*}[!t]
\centering
\begin{tikzpicture}
\begin{axis}[
    width=0.75\textwidth,
    height=4.8cm,
    xbar,
    xmin=0, xmax=125,
    xtick={0,25,50,75,100,125},
    ytick=data,
    y dir=reverse,
    bar width=8pt,
    enlarge y limits=0.22,
    tick label style={font=\footnotesize},
    yticklabel style={font=\footnotesize, align=right},
    axis x line*=bottom,
    axis y line*=left,
    grid=major,
    major grid style={draw=black!10},
    clip=true,                 
    enlarge x limits={upper, value=0.06},
    yticklabels={
        Algorithms \& Data Structures,
        Frontend Development,
        Backend Development,
        Database \& Querying,
        DevOps \& Infrastructure,
        Security \& Vulnerability Manag\ldots,
        Data Processing \& Scripting
    },
]

\addplot[fill=black!35] coordinates {
    (123,0)
    (87,1)
    (80,2)
    (75,3)
    (18,4)
    (21,5)
    (41,6)
};

\node[font=\scriptsize, anchor=center] at (axis cs:61.5,0) {123 (78.3\%)};
\node[font=\scriptsize, anchor=center] at (axis cs:43.5,1) {87 (55.4\%)};
\node[font=\scriptsize, anchor=center] at (axis cs:40.0,2) {80 (51.0\%)};
\node[font=\scriptsize, anchor=center] at (axis cs:37.5,3) {75 (47.8\%)};
\node[font=\scriptsize, anchor=center] at (axis cs:20.5,6) {41 (26.1\%)};

\node[font=\scriptsize, anchor=west] at (axis cs:18.8,4) {18 (11.5\%)};
\node[font=\scriptsize, anchor=west] at (axis cs:21.8,5) {21 (13.4\%)};

\end{axis}
\end{tikzpicture}
\caption{Primary technical domains for which respondents utilize LLM assistance (multi-select; $N=157$).}
\label{fig1}\vspace{-3mm}
\end{figure*}
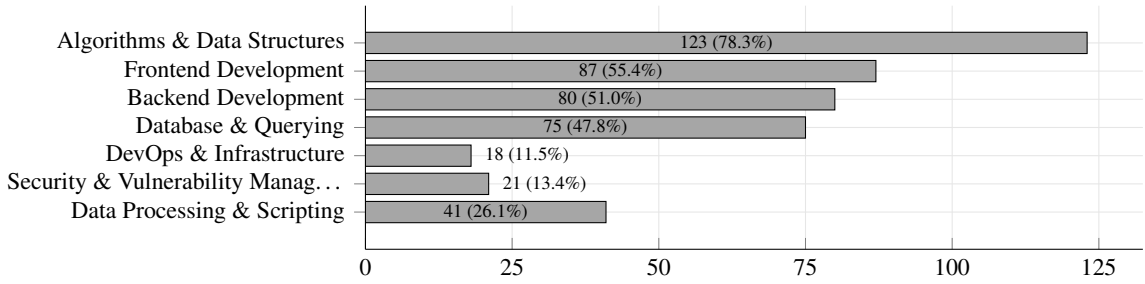

\paragraph{Where respondents perceive LLMs to perform poorly.\vspace{-2mm}}
Respondents most frequently reported that LLMs perform poorly when debugging complex logic errors in large projects (54.78\%, $86/157$). This is followed by connecting multiple components (16.56\%, $26/157$) and solving LeetCode-style algorithmic problems (14.01\%, $22/157$). This is noteworthy as it contrasts with the domain-use patterns presented earlier. For example, although respondents \emph{use} LLMs heavily for algorithms and data structures (Figure \ref{fig1}), many still identify algorithmic problem solving as a weakness.

\begin{table*}[t]
\centering
\footnotesize
\caption{Scenario judgments (collapsed to winner/equality; $N=157$, 1Y=$131$, 4Y=$26$).}
\label{tab:scenario_judgments}
\begin{tabular}{llrrrrrr}
\toprule
\textbf{Scenario} & \textbf{Judgment} & \textbf{\textit{N}} & \textbf{\%} & \textbf{\textit{1Y}} & \textbf{\%} & \textbf{\textit{4Y}} & \textbf{\%} \\
\midrule

\textbf{Ambiguous interpretation} 
& Human domain expert & 92 & 58.60 & 74 & 56.49 & 18 & 69.23 \\
& LLM & 45 & 28.66 & 39 & 29.77 & 6 & 23.08 \\
& Both/equal & 20 & 12.74 & 18 & 13.74 & 2 & 7.69 \\
\addlinespace

\textbf{Speed initial draft} 
& LLM & 132 & 84.08 & 108 & 82.44 & 24 & 92.31 \\
& Both/equal & 14 & 8.92 & 13 & 9.92 & 1 & 3.85 \\
& Human domain expert & 11 & 7.01 & 10 & 7.63 & 1 & 3.85 \\
\addlinespace

\textbf{Domain rules/constraints} 
& Human domain expert & 82 & 52.23 & 61 & 46.56 & 21 & 80.77 \\
& LLM & 47 & 29.94 & 45 & 34.35 & 2 & 7.69 \\
& Both/equal & 28 & 17.83 & 25 & 19.08 & 3 & 11.54 \\
\addlinespace

\textbf{Iterative convergence} 
& LLM & 68 & 43.31 & 56 & 42.75 & 12 & 46.15 \\
& Human domain expert & 55 & 35.03 & 46 & 35.11 & 9 & 34.62 \\
& Both/equal & 34 & 21.66 & 29 & 22.14 & 5 & 19.23 \\
\bottomrule
\end{tabular}\vspace{-4mm}
\end{table*}

\subsubsection{Scenario judgments: perceived advantage of humans vs.\ LLMs\vspace{-2mm}}
Table~\ref{tab:scenario_judgments} reports scenario judgments collapsed to (i) LLM advantage, (ii) human expert advantage, or (iii) parity. For tasks involving unclear instructions and interpretation, both cohorts favored the human expert (58.60\% overall), with a stronger human preference among Fourth-Year students (69.23\%) than First-Year students (56.49\%). Under a 30-minute constraint, both cohorts overwhelmingly favored LLMs (84.08\% overall). Fourth-Year respondents were even more likely to prefer the LLM for speed (92.31\%). This suggests that with increased experience, students increasingly conceptualize LLMs as high-throughput drafting tools.

The largest cohort difference appears for tasks involving unspoken domain rules and constraints. Overall, a majority favored the human expert (52.23\%), but Fourth-Year respondents were much more likely to prefer humans (80.77\%) than First-Year respondents (46.56\%), who were split between human advantage, LLM advantage, and parity. This difference is consistent with a developmental progression in which increased experience (and especially industry exposure) increases awareness of compliance requirements, invariants, and latent constraints that are not captured by surface-level plausibility.

For iterative refinement, views were more balanced. 43.31\% favored LLMs, 35.03\% favored humans, and 21.66\% perceived parity. Notably, both cohorts show a similar split, suggesting that students see dialogue-based iteration as a setting where either agent can be effective depending on the quality of feedback and the nature of the task.

\subsubsection{Between-cohort differences and within-sample associations}
Because the Fourth-Year subgroup is small ($n=26$) relative to First-Year ($n=131$), comparisons should be interpreted as suggestive. Following empirical guidance for survey studies, we emphasize nonparametric tests for ordinal variables and report effect sizes alongside p-values (Table~\ref{tab:between_year_differences}).

Fourth-Year respondents reported substantially higher programming proficiency (median 4.0 vs.\ 2.0; Mann--Whitney $U=261.0$, $p<0.001$, rank-biserial $r=0.85$) and higher confidence without external help (median 4.0 vs.\ 3.0; $U=1106.0$, $p=0.002$, $r=0.35$). They also reported longer LLM usage histories (median category 2.0 vs.\ 1.0; $U=767.5$, $p<0.001$, $r=0.55$).
Daily-workflow LLM use was more prevalent among Fourth-Year students (84.62\%) than First-Year students (50.38\%; Fisher's exact $p=0.0011$, OR$=5.42$), indicating substantially higher routine integration in the senior cohort.

A particularly strong cohort difference emerged in the domain-rules scenario. Here, Fourth-Year students showed a stronger human preference (median 2.0 vs.\ 0.0; $U=839.0$, $p<0.001$, $r=0.51$). To further examine this relationship, an exploratory analysis was conducted within the First-Year subgroup. Even within this cohort, students with industry experience exhibited a significantly stronger preference for human-generated solutions in constraint-heavy tasks ($U = 612.5$, $p = 0.029$, $r = 0.33$). This suggests that industry exposure may change how students judge “correctness.” Instead of focusing only on whether the code looks syntactically correct, experienced students may pay more attention to whether it truly respects real-world constraints, etc.
 
Several within-sample associations contextualize reliance, trust, and skill. For example, confidence without external help was negatively associated with reliance on LLMs for assignments (Spearman $\rho=-0.19$, $p=0.017$), indicating that less confident respondents rely more on LLMs for assessed tasks. Longer reported history of LLM usage was weakly associated with more frequent workflow use ($\rho=0.19$, $p=0.016$), consistent with habituation effects.
Programming proficiency showed two complementary associations. Here, higher proficiency was associated with stronger preference for \emph{humans} in constraint-heavy tasks (domain rules scenario; $\rho=0.35$, $p<0.001$), while also being associated with stronger preference for \emph{LLMs} in speed-focused drafting (speed scenario; $\rho=-0.30$, $p<0.001$). This pairing suggests that more proficient respondents may simultaneously recognize (i) LLMs' advantage in rapid first-pass generation and (ii) humans' advantage when correctness depends on implicit constraints and compliance requirements.

Finally, verification effort increased slightly with programming proficiency ($\rho=0.21$, $p=0.009$). Combined with the negative association between perceived reliability and verification effort ($\rho=-0.21$, $p=0.008$), this indicates that skill may function as a protective factor: more proficient respondents both (i) apply stricter validation standards and (ii) are less likely to reduce oversight simply because outputs \emph{seem} to work.

\setlength{\tabcolsep}{5pt}
\begin{table*}[t]
\centering
\footnotesize
\caption{Selected between-year differences (effect-size oriented reporting). Positive $r$ indicates higher values in Fourth-Year.}
\label{tab:between_year_differences}
\begin{tabular}{p{5.75cm}p{1.9cm}p{1.9cm}p{1.6cm}p{1.2cm}p{1.2cm}}
\toprule
\textbf{Measure} & \textbf{First-Year median [IQR]} & \textbf{Fourth-Year median [IQR]} & \textbf{Test stat} & \textbf{\textit{p}} & \textbf{Effect size} \\
\midrule
Self-rated programming proficiency (ordinal) 
& 2.0 [1.5, 3.0] 
& 4.0 [3.25, 4.0] 
& $U = 261.0$ 
& $< 0.001$ 
& $r = 0.85$ \\
Confidence without external help (ordinal) 
& 3.0 [3.0, 3.0] 
& 4.0 [3.0, 4.0] 
& $U = 1106.0$ 
& $= 0.002$ 
& $r = 0.35$ \\
Years of LLM use (ordinal) 
& 1.0 [1.0, 2.0] 
& 2.0 [2.0, 2.75] 
& $U = 767.5$ 
& $< 0.001$ 
& $r = 0.55$ \\
Number of domains using LLMs (count; 1--7) 
& 2.0 [1.0, 3.0] 
& 4.5 [3.0, 5.0] 
& $U = 603.0$ 
& $< 0.001$ 
& $r = 0.65$ \\
Domain rules sc. (human advantage; ordinal) 
& 0.0 [$-1.0$, 2.0] 
& 2.0 [1.0, 3.0] 
& $U = 839.0$ 
& $< 0.001$ 
& $r = 0.51$ \\
\bottomrule
\end{tabular}\vspace{-3mm}
\end{table*}

\subsubsection{Implications from the Survey results}
First, LLM use has become embedded in the everyday workflows of most students, particularly senior cohorts. Consequently, research designs that assume a clearly defined “no-AI” baseline may lack ecological validity. Rather than conceptualizing LLM use as a simple binary condition (use vs. non-use), future studies should account for variation in adoption intensity.

Second, multiple indicators point to a \emph{trust-calibration gradient}. Novices are more likely to treat LLMs as capable in contexts where implicit constraints matter (e.g., domain rules) and more likely to reduce verification under high perceived reliability. However, more experienced users both verify more and reserve humans for constraint-heavy correctness. This motivates measuring not only \emph{frequency} of use, but also \emph{verification depth}, \emph{prompting/iteration strategies}, and \emph{error-detection competence} in follow-up work.\vspace{-2mm}

\begin{table*}[t!]
\centering
\footnotesize
\caption{Theme--code map (open coding) with contributing participants. (4th year in Bold)}
\label{tab:results_codes}
\begin{tabularx}{\textwidth}{>{\raggedright\arraybackslash}p{3.0cm} >{\raggedright\arraybackslash}X >{\raggedright\arraybackslash}p{3.2cm}}
\toprule
\textbf{Theme} & \textbf{Open code (definition)} & \textbf{Contributing participants} \\
\midrule

Task framing \& triage
& \textbf{AI-first entry}: prompting precedes visible independent decomposition (no pseudocode/notes before AI). & S02, S03, S06, S10, S14, \textbf{S09, S15, S16, S18} \\
& \textbf{Deliberate requirement internalization}: slower reading and constraint review before tooling/coding. & S01, S04, S05, S07, S13, \textbf{S11, S17, S19} \\
& \textbf{Breadth-first switching}: frequent jumps across sections before resolving earlier tasks. & S01, S02, S03, S06, S10 \\
& \textbf{Low-effort entry-point}: beginning with MCQ/pattern responses prior to coding work. & S07 \\

\addlinespace
Knowledge sourcing \& planning
& \textbf{LLM as compressed documentation}: framework/tool learning performed via chat rather than external references. & S02, S06, \textbf{S09} \\
& \textbf{Official documentation confirmation}: use of authoritative docs to validate setup patterns before coding. & \textbf{S11} \\
& \textbf{Example-driven web planning}: reliance on tutorials/StackOverflow templates to plan and implement. & S01, S05, S07 \\

\addlinespace
Solution construction \& reuse
& \textbf{Copy--transfer integration}: AI output pasted into IDE with minimal restructuring. & S02, S06, S10, S14, \textbf{S15, S16, S18} \\
& \textbf{AI as procedural instructor}: stepwise checklist-following for setup and implementation. & S06, S14, \textbf{S15, S16} \\
& \textbf{Multi-LLM switching}: moving across models to overcome limits or seek improved outputs. & S02, S06, \textbf{S09, S18} \\
& \textbf{Non-LLM template reuse}: reuse of prior repos/projects/local notes as scaffolds. & \textbf{S09, S12, S17, S19} \\

\addlinespace
Debugging \& verification
& \textbf{AI-mediated debugging}: errors pasted into LLM; fixes enacted procedurally. & S02, S03, S06, S14, \textbf{S09, S15, S16} \\
& \textbf{Reactive trial--error loops}: compile/run $\rightarrow$ search $\rightarrow$ patch cycles without AI. & S04, S05, S07, \textbf{S11, S17, S19, S18} \\
& \textbf{Execution-based validation}: explicit runs/tests (API calls, test cases) to check correctness. & S03, S05, S07, \textbf{S09}, S11, \textbf{S12 S15, S17, S19} \\
& \textbf{Cross-tool persistence verification}: API testing plus DB inspection as proof-of-save. & \textbf{S09, S17, S19} \\
& \textbf{Low verification acceptance}: scrolling/reviewing without running or edge-case checking. & S02, S06, S10, S14, \textbf{S16} \\

\addlinespace
Toolchain \& environment management
& \textbf{Dependency boundary stalls}: missing runtimes/libraries impede completion despite available code. & S02, S03, S06, S01, S04 \\
& \textbf{AI-guided environment recovery}: installation/configuration guided via LLM instructions. & S06, S14, \textbf{S09} \\
& \textbf{CLI-centered toolchain fluency}: conventional setup/testing via venv, pip, curl, DB CLI, etc. & \textbf{S11, S17, S19} \\

\addlinespace
Architecture \& explanatory work
& \textbf{LLM-authored architecture answers}: design patterns and trade-offs primarily generated by AI. & S02, S06, S10, S14, \textbf{S09, S15, S16, S18} \\
& \textbf{Participant-authored modeling}: diagrams/notes/tables used to externalize design reasoning. & \textbf{S17, S19, S20} \\
& \textbf{Operational risk framing}: emphasis on scaling risks, failure modes, and system behavior under load. & \textbf{S09}, S17, S19 \\
& \textbf{AI-assisted pattern reasoning}: screenshot/image prompting used to answer visual/pattern tasks. & S02, S14, \textbf{S16, S18} \\
\bottomrule
\end{tabularx}
\vspace{-2mm}
\end{table*}

\subsection{Task-Based Quasi-Experiment}
 The findings of this experiment are presented below and summarized in Table~\ref{tab:results_codes}.

\subsubsection{Task framing and triage}

AI assistance changed how participants entered tasks. In the AI conditions, task entry was frequently ``AI-first''. For example, several first-year participants opened the PDF and prompted an LLM for full solutions before producing visible independent decomposition artifacts (e.g., pseudocode, constraint lists) [S02, S06, S10, S14]. More specifically, [S06] prompted Claude for a complete REST API solution within the first minutes of viewing the task sheet, delegating the choice of stack (Node/Express/MongoDB) [S06]. Similarly, [S14] copied multiple Section~01 questions into ChatGPT before drafting a plan and repeatedly requested ``clear'' answers as an answer-sheet artifact [S14]. A comparable entry pattern appeared among fourth-year AI users. However, prompt tailoring sometimes reflected pre-existing preferences (e.g., specifying Spring Boot/PostgreSQL) even when implementation details were delegated [S09, S15, S16, S18].

Without AI, triage more often began with deliberate reading and internalization of requirements prior to tool selection. First-year participants in the non-AI condition tended to scroll more slowly, re-check constraints, and only then move to searching or coding, indicating heavier reliance on internal mental modeling even when incomplete [S01, S04, S05, S13]. Fourth-years without AI showed the clearest evidence of deliberate triage, including early decomposition and structured task sequencing [S11, S17, S19]. For instance, [S11] read the task set and then consulted official Flask and Flask-PyMongo documentation before writing the route structure, reflecting a verification-oriented planning phase [S11]. In the absence of AI, task switching was often triggered by blockage (e.g., stalling on REST API feasibility and pivoting to simpler tasks) rather than by low-cost AI-enabled parallelism [S01, S03]. One non-AI first-year S07 prioritized lower-effort selection tasks (pattern/MCQ answers) prior to bounded coding tasks, suggesting time-management behavior under load [S07].

\subsubsection{Knowledge sourcing and planning: LLMs versus documentation}
In AI-assisted workflows, LLMs frequently served as a compressed substitute for documentation and conceptual scaffolding. Participants used conversation to learn unfamiliar frameworks or constraints (e.g., asking what \texttt{crow.h} is, or why a particular sorting algorithm minimizes writes) rather than navigating official references [S02, S06]. S02 uploaded the PDF to Gemini, accepted a Crow-based REST API proposal, and then asked Gemini to explain the framework [S02]. S06 relied on Claude both for stack selection and for environment setup guidance when Node was not recognized in the shell, indicating planning-through-instructions rather than planning-through-reference [S06].

In the non-AI condition, planning relied more heavily on the open web (tutorial sites, StackOverflow) and, in the fourth-year cohort, on official documentation. A first-year without AI [S05] repeatedly searched for language semantics (e.g., \texttt{strcmp}, Python list mutation) and algorithm templates, reflecting example-driven planning that was reactive to errors and performance constraints [S05]. By contrast, S11 (fourth-year without AI) used authoritative sources (Flask and Flask-PyMongo docs) to confirm setup patterns prior to implementation, reflecting an ``expert verification'' approach [S11]. This contrast suggests that AI shifted planning upstream into prompting, whereas non-AI workflows shifted planning into browsing and incremental confirmation [S02, S06, S11, S05].

\subsubsection{Solution construction and reuse strategies}
A dominant construction pattern in AI-assisted sessions was ``copy--transfer'': the LLM generated a near-complete template, and the participant enacted integration steps (creating directories, installing dependencies, pasting code) with minimal structural modification [S06, S10, S14]. S10 integrated AI-generated Node/Express/Mongoose scaffolding for the REST API and moved across tasks by copying each prompt verbatim, often scrolling quickly through responses rather than re-deriving logic [S10]. S14 followed ChatGPT step-by-step for environment setup and repeatedly reprompted using compile errors, converging via AI-directed procedural correction rather than independent debugging [S14]. Within fourth-year AI sessions, similar copy--transfer behavior was observed, particularly under time pressure or with explanation-style questions, where answers were copied or rephrased without deep interrogation [S15, S16, S18].

Two additional reuse modes differentiated cohorts. First, multi-LLM switching emerged as an AI-specific strategy: participants changed models to overcome limits, seek more executable outputs, or improve completeness [S02, S06, S09, S18]. For example, S02 moved from Gemini to DeepSeek to obtain an expanded REST API implementation and later switched tools again for pattern-related tasks [S02]. Second, non-LLM template reuse was prominent among senior non-AI participants: fourth-years without AI often reused prior project structures, local notes, and conventional scaffolds rather than browsing for ad hoc snippets [S12, S17, S19]. This contrasts with first-year non-AI participants, who tended to intermingle multiple partial implementations in a single workspace and relied on web examples for each isolated subproblem [S01, S04, S07].

\subsubsection{Debugging and verification behavior}
Verification practices varied sharply by condition and cohort. In first-year AI sessions, a recurrent pattern was low verification intensity: solutions were frequently accepted after cursory review, with limited evidence of test-case execution for complex tasks (especially architecture/explanatory prompts) [S10, S06]. For example, S10 reviewed multi-language algorithm answers and architecture outputs largely by scrolling and then advanced without rerunning or independently validating edge cases [S10]. Similarly, S06 switched between Claude and Gemini to obtain additional answers while leaving some implementations incomplete due to unresolved environment constraints [S06].

When debugging did occur in AI sessions, it was often AI-mediated: participants pasted raw error messages into an LLM and followed procedural fix instructions [S02, S03, S14, S16]. S03 used AI to interpret a C compilation constraint (non-constant array sizing), adopted the suggested memory allocation fix, and progressed quickly after the AI diagnosis [S03]. S14 repeatedly provided compile errors to ChatGPT and converged via AI-directed adjustments, indicating an ``LLM as debugger'' workflow [S14]. Among fourth-year AI participants, the same mechanism was present but ranged from calibrated use to blind compliance; S16 was described as repeatedly copying errors into ChatGPT and following instructions without evident internal diagnosis [S16].

In non-AI sessions, debugging was more often characterized by reactive trial--error loops supported by web search (first-years) or by systematic test-and-verify routines (fourth-years). S07 adapted a StackOverflow bracket-validation template and iteratively corrected errors through compile/run cycles and additional searches, showing incremental refinement and execution-based validation [S07]. S05 demonstrated frequent execution attempts and targeted searches, but also displayed conceptual misattribution in simulation reasoning (investigating argument passing rather than isolating in-place mutation) [S05]. In contrast, fourth-years without AI displayed stronger verification discipline, including API testing and database inspection as proof-of-persistence artifacts [S17, S19], and in S11’s case, a documentation-confirmed environment setup before implementation [S11]. These patterns suggest that AI reduced the cost of obtaining candidate fixes but did not consistently increase verification rigor; verification remained coupled to participants' baseline practices and tooling fluency [S10, S16, S17, S19].

\subsubsection{Toolchain and environment management}
Environmental constraints repeatedly shaped outcomes and exposed an important limit of AI scaffolding: generated solutions frequently assumed dependencies that were not available in the execution context. Several AI-assisted first-years attempted advanced stacks that failed due to missing runtimes or libraries (e.g., C++ Crow headers not present, Node not installed or not on PATH), leading to stalled progress despite having ``complete'' generated code [S02, S03, S06]. S02 encountered linker and missing-header errors when attempting Crow-based code in online sandboxes and then sought AI explanations for toolchain limitations, illustrating a dependency boundary that the LLM could describe but not resolve within the constrained environment [S02]. S06 encountered a missing Node runtime and transitioned into AI-guided installation steps rather than diagnosing the environment independently [S06].

Non-AI first-years also experienced toolchain friction (directory/path errors, IDE project-creation issues), but without AI the recovery pathways were slower and sometimes incomplete by session end [S01, S04]. Fourth-years showed stronger environment navigation overall, but AI could still redirect attention away from root-cause debugging toward workarounds. S09 attempted a CLI database command, encountered a missing client, and then shifted to a GUI pathway (DBeaver) after consulting ChatGPT for alternate steps, demonstrating adaptive tool substitution rather than extended diagnosis [S09]. Fourth-years without AI more consistently relied on conventional environment construction and verification steps (e.g., virtual environments, CLI testing, structured scaffolds), suggesting a more stable toolchain schema independent of AI assistance [S11, S17, S19].

\subsubsection{Architecture and explanatory tasks}
The largest qualitative contrast appeared in open-ended architecture and explanation tasks. In AI-assisted sessions, architecture work frequently took the form of LLM-authored answer production: participants copied scenario text into the model and received pattern-rich responses (API Gateway, adapters, canonical models, event-driven backbones), which were then consumed as final answers with limited participant-authored modeling artifacts [S02, S06, S10, S14]. For example, S10 prompted middleware questions verbatim and advanced by rapidly reviewing AI responses without producing independent diagrams or trade-off matrices [S10]. S14 requested ``clear'' solutions and copied outputs into a document, treating the LLM response as an answer sheet [S14]. Similar behavior was reported among fourth-year AI participants for explanation-driven questions, including rephrasing via another model rather than reconstructing reasoning [S15, S16].

Without AI, first-years often deferred architecture tasks or engaged via narrative reasoning rather than structured models, indicating lower confidence in open-ended abstraction tasks under time constraints [S01, S04, S07, S13]. Fourth-years without AI were distinct in producing structured design artifacts (notes, diagrams/tables, operational reasoning), including explicit risk framing (throughput, backpressure, idempotency, failure modes) as part of architectural justification [S17, S19]. Notably, when AI was paired with strong senior practices, AI outputs could be integrated into a verification-centered workflow: S09 used AI to scaffold architectural and implementation steps but still validated the REST endpoint through API testing and confirmed persistence via database inspection, indicating that AI served as an accelerator rather than a substitute for end-to-end reasoning [S09]. Conversely, AI also supported non-coding reasoning tasks (pattern puzzles) via screenshot prompting, sometimes with no manual attempt, highlighting that AI influenced general reasoning behaviors in addition to programming behaviors [S02, S14, S16, S18].

\subsubsection{Implications of the experiment results}
The findings suggest that AI assistance should be treated as a spectrum of practice rather than a binary condition, because AI use ranged from calibrated augmentation (i.e., AI as accelerator with verification) to procedural substitution (i.e., AI as answer sheet with minimal reading) even within the same academic year [S09, S16, S18]. In pedagogy, this motivates explicit instruction in \emph{AI literacy} as a verification-centered discipline. Here, students should be required to produce verification artifacts (e.g., test cases, API requests, database evidence, or minimal reproducible cases) alongside code and explanations, given that these practices could distinguish rigorous workflows from superficial task completion under both AI-assisted and unassisted conditions [S03, S09, S17, S19]. 

For assessment design, the results imply that grading solely on final outputs may overestimate competence when AI-generated scaffolds are integrated without participant-owned reasoning; process evidence (tests, trade-off rationales, and environment setup documentation) should therefore be incorporated to evaluate conceptual understanding [S10, S16, S17]. Additionally, repeated environment failures indicate that toolchain competence remains a bottleneck that AI does not reliably solve. Here, curricula should emphasize dependency management, runtime configuration, and debugging hierarchies as foundational skills that enable students to realize (or critically reject) generated solutions [S02, S06, S14].\vspace{-4mm}

\section{\uppercase{Threats to Validity}}
\label{sec:threats}

\vspace{-2mm}\textbf{Internal validity.} Phase~2 is quasi-experimental and uses purposive grouping based on prior LLM experience and stated preferences. Consequently, group differences may reflect pre-existing factors (e.g., proficiency, tool fluency, motivation) in addition to AI availability. \textbf{Construct validity.} Several constructs (e.g., ``AI literacy,'' verification effort, and confidence) are measured via self-report and may not fully match observed behavior. Task performance was done through a bounded task suite. While it spans multiple abstraction levels, it cannot represent the full breadth of industrial SE (e.g., long-term maintenance and team coordination). Video transcription and timeline construction of the experiments were done using tools/services, and this may introduce potential errors. \textbf{External and conclusion validity.} Participants are undergraduates from a single institution, and the quasi-experiment sample is small, which limits generalizability.\vspace{-4mm}

\section{\uppercase{Conclusion}\vspace{-3mm}}
\label{sec:conclusion}

This study examined LLM assistance in software engineering using a mixed-methods design that combined a survey and a task-based quasi-experiment. The survey shows that LLM use is already routine among students, but trust calibration and verification practices vary with seniority and experience. The quasi-experiment indicates that AI availability reshapes workflow strategies, often shifting effort from upfront planning and documentation navigation toward prompting, integration, and AI-mediated debugging. Collectively, we believe that our findings temper broad claims that LLMs uniformly ``bridge'' expertise gaps. For example, LLMs can accelerate drafting and reduce friction for routine implementation, but they do not remove the need for human judgment tasks such as constraint interpretation, toolchain troubleshooting, and validation. However, future work should further validate these findings with larger and more diverse populations and incorporate richer process measures (e.g., time allocation, test artifacts, and debugging episodes) to better explain when AI assistance helps and when it merely relocates effort.\vspace{-4mm}

\section*{Acknowledgements}

This work was supported by the University of Colombo School of Computing (UCSC) Research Allocation for Research and Development.\vspace{-4mm}

\bibliographystyle{apalike}
{\small
\bibliography{example}}

\end{document}